\documentclass[preprint,preprintnumbers,nofootinbib,aps,10pt,twocolumn]{revtex4-1}
\usepackage{amsmath,amssymb,bm,epsfig}
\usepackage{color}
\usepackage{natbib}
\usepackage{hyperref} 
\usepackage{ulem}
\usepackage{graphicx}
\usepackage{xcolor,colortbl}
\RequirePackage{lineno}
\newcommand{\nn}{\nonumber}
\newcommand{\sNN}{\sqrt{s_{\textrm{NN}}}}

\definecolor{Gray}{gray}{0.85}
\newcolumntype{a}{>{\columncolor{Gray}}c}

\def \beq{\begin{equation}}
\def \eeq{\end{equation}}
\def \beqa{\begin{eqnarray}}
\def \eeqa{\end{eqnarray}}
\def \la{\langle}
\def \ra{\rangle}
\def \l{\left(}
\def \r{\right)}
\def \l{\left(}
\def \r{\right)}

\begin{document}

\title{Large directed flow of open charm mesons  probes the  three dimensional distribution of matter in  heavy ion collisions}

\author{Sandeep Chatterjee}
\email{Sandeep.Chatterjee@fis.agh.edu.pl}
\affiliation{AGH University of Science and Technology,\\ 
Faculty of Physics and Applied Computer Science,\\
aleja Mickiewicza 30, 30-059 Krakow, Poland}

\author{Piotr Bo{\.z}ek}
\email{Piotr.Bozek@fis.agh.edu.pl}
\affiliation{AGH University of Science and Technology,\\ 
Faculty of Physics and Applied Computer Science,\\
aleja Mickiewicza 30, 30-059 Krakow, Poland}

\begin{abstract}

Thermalized matter  created in non-central
 relativistic heavy-ion collisions is expected to be 
tilted in the reaction plane with respect 
to the beam axis. 
The most notable consequence of this forward-backward symmetry 
breaking is the observation of 
rapidity-odd directed flow for charged particles. 
On the other hand, the production points for  heavy quarks  
are forward-backward symmetric and shifted
 in the transverse plane with respect to the fireball. The drag on heavy quarks from the asymmetrically 
distributed thermalized matter generates substantial directed flow 
for heavy flavor mesons.
We predict a very large rapidity odd directed flow of $D$ mesons in non-central Au-Au collisions at $\sqrt{s_{NN}}=200$~GeV, {\it several times larger} 
than for charged particles. A possible experimental observation of a large directed flow for
heavy flavor mesons  would represent an almost direct probe of the 3-dimensional distribution of matter in heavy-ion collisions.

\end{abstract}

\maketitle

The dynamics of relativistic heavy-ion collisions can be largely factorized into a part that is transverse to the 
beam direction and the longitudinal dynamics that is parallel to it. The transverse evolution has been 
extensively investigated with ample evidence of collective
 expansion~\cite{Ollitrault:2010tn,*Heinz:2013th,*Gale:2013da}. 
The longitudinal dynamics is usually studied in two aspects, 
 event by event fluctuations of the longitudinal
distribution of matter   \cite{Armesto:2006bv,*Bzdak:2009xq,*Bozek:2010vz, *Petersen:2011fp,*Bzdak:2012tp, *Jia:2014vja,*Pang:2014pxa,*Monnai:2015sca} and 
 the breaking of the beam axis symmetry in non-central collisions
 \cite{Csernai:1999nf,*Snellings:1999bt,*Lisa:2000ip,*Becattini:2013vja,Bozek:2010bi}.
 One of the key question towards understanding the longitudinal 
evolution is how  the forward-backward symmetry is broken in the initial state  
for symmetric but non-central collision systems. 
 The determination 
of the strength and mechanism of this symmetry breaking will lead
 to a significant progress of our understanding of 
energy deposition in relativistic collisions.

\begin{figure}
\begin{center}
\hskip -3mm
 \includegraphics[scale=0.4]{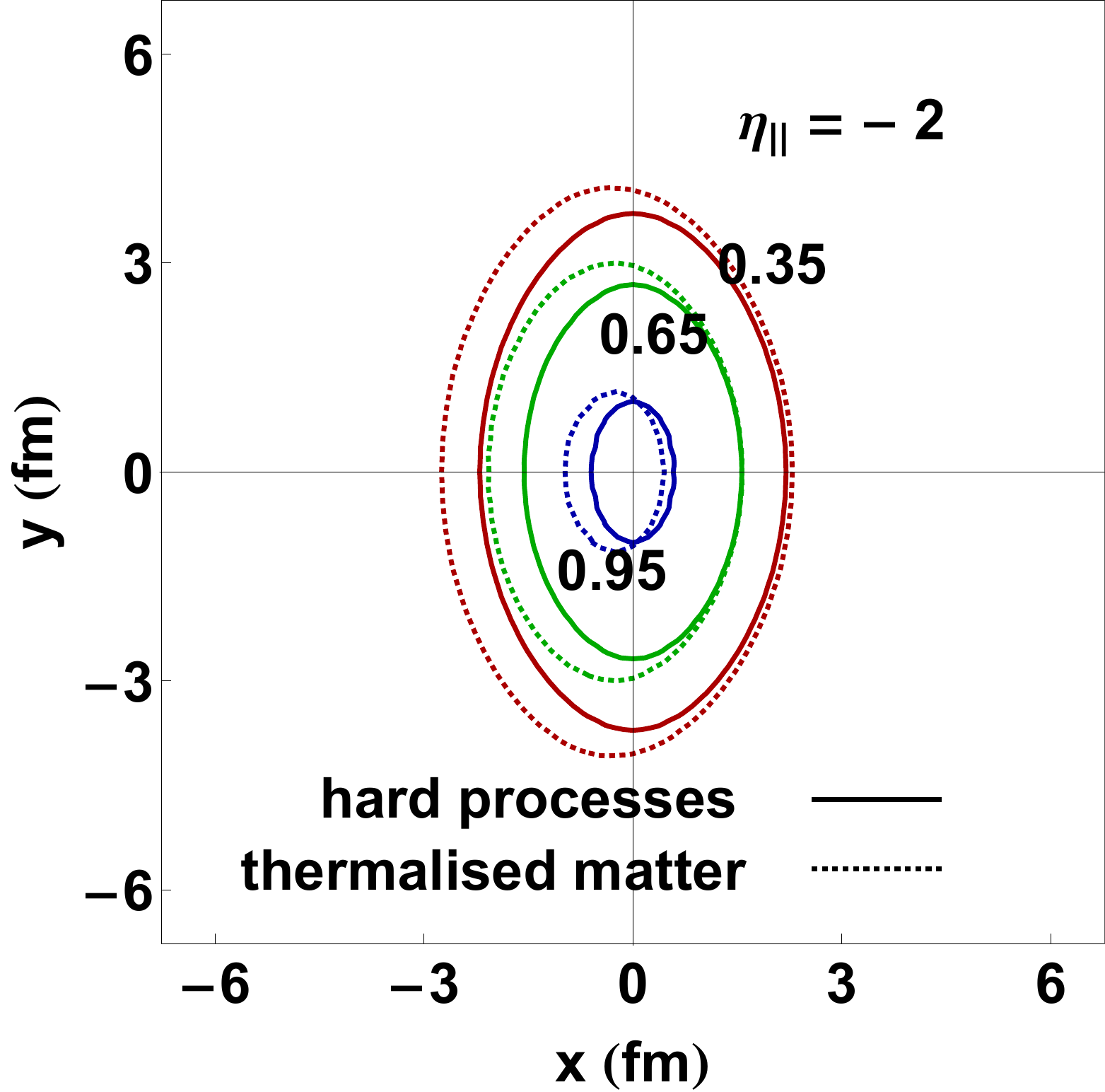}
 \caption{ The transverse density  profile of the fireball tilted relative to the beam axis  and the binary collision profile at $\eta_{||}=-2$. }
 \label{fig.tilt}
\end{center}
\vskip -5mm

\end{figure}

At intermediate beam energies, charged particle directed flow $v_1$ is sensitive to the QCD equation of 
state~\cite{Steinheimer:2014pfa,*Nara:2016hbg}. At higer energies ($\sqrt{s_{NN}}=200$GeV and above), it is a key observable that 
probes the longitudinal profile of the fireball~\cite{Bozek:2010bi}. Heavy quarks with masses that are several times larger than the fireball temperature are 
mostly produced in the initial stages of the collision. Hence they witness the entire space-time evolution of the fireball 
and are ideal candidates to probe initial state physics~\cite{Rapp:2009my,*Andronic:2015wma,*Aarts:2016hap,*Noronha-Hostler:2016eow}. We demonstrate 
the possibility of probing the initial state longitudinal profile of the thermalized matter through the directed flow of heavy quarks.

A natural suggestion to explain the rapidity spectra of produced particles in asymmetric collisions like d+Au~\cite{Nouicer:2004ke} 
is that each participant source preferably showers particles along its direction of motion~\cite{Bialas:2004su, Bzdak:2009xq}. 
Such a picture is also well motivated by the parton model~\cite{Brodsky:1977de} as well as saturation models of 
QCD~\cite{Adil:2005qn}. In order to obtain such a final state within a hydrodynamic framework, the following ansatz for 
the initial entropy density $s\l\tau_0,x,y,\eta_{||}\r$ within a two-component Glauber model has been proposed~\cite{Bozek:2010bi}
\beqa
 s\l\tau_0,x,y,\eta_{||}\r &=& s_0 \left[ \alpha N_{coll}+\l1-\alpha\r\l N_{part}^+ f_+\l\eta_{||}\r+\right.\right.\nn\\
 && \left.\left.N_{part}^- f_-\l\eta_{||}\r\r\right]f\l\eta_{||}\r\label{eq.ic}
\eeqa
where $N_{coll}$, $N_{part}^+$ and $N_{part}^-$ are the densities for number of binary collisions and participants from 
the forward and backward going nuclei respectively evaluated at $\l x,y\r$. $\tau=\sqrt{t^2-z^2}$ and 
$\eta_{||}=\frac{1}{2}\log\frac{\l t+z\r}{\l t-z\r}$ are the proper time and spacetime rapidities respectively. $\tau_0$ is 
the initial proper time to start hydrodynamic evolution. 
\beqa
 f\l\eta_{||}\r &=& \exp\l-\theta\l|\eta_{||}|-\eta_{||}^0\r\frac{\l|\eta_{||}|-\eta^0_{||}\r^2}{2\sigma^2}\r 
 \label{eq.feta}
\eeqa
 is the longitudinal profile (for $|\eta_\parallel|<y_{beam}$) taken  to 
reproduce the final state rapidity spectra of charged particles.
$\alpha$ controls the admixture of the participant and binary sources contribution to the total entropy deposition for the fireball. 
For a good description of the Au+Au data at $\sNN=200$ GeV, we set $\eta^0_{||}=1.3$, $\sigma=1.5$ and $\alpha=0.15$. 

$f_{+,-}\l\eta_{||}\r$ are the profiles which introduce the rapidity 
odd component to the initial state
\beq
 f_+\l\eta_{||}\r = \left\{\begin{array}{lr}
                            0, & \eta_{||}<-\eta_T\\
                            \frac{\eta_T+\eta_{||}}{2\eta_T}, & -\eta_T\leq\eta_{||}\leq\eta_T\\
                            1, & \eta_{||}>\eta_T
                           \end{array}\right.
\label{eq.fpm}
\eeq
with  $f_-\l\eta_{||}\r=f_+\l-\eta_{||}\r$. 
The above tilted initial condition has been 
shown to describe the rapidity odd $v_1$ of charged 
particles~\cite{Bozek:2010bi} measured by STAR~\cite{Abelev:2008jga} and 
PHOBOS~\cite{Back:2005pc}. The magnitude of the tilt of the initial source 
is determined by the parameter $\eta_T$. Typically we use  
$\eta_T\simeq y_{beam}-2=3.36$, the value extracted  from 
d+Au collisions \cite{Bialas:2004su}, but
to demonstrate the great sensitivity of heavy flavor flow to initial tilt we
will  also vary $\eta_T$  between $2$ and the beam rapidity $y_{beam}=5.36$.

The transverse profile of the entropy density 
for the fireball and of the binary 
collision sources at $\eta_{\parallel}=-2$ for Au-Au collision
 at impact parameter $b=8.3$~fm is plotted 
in Fig.~\ref{fig.tilt}. The Au nucleus flying with the positive (negative) 
rapidity is located at $x=b/2(-b/2)$.
 The fireball
 is tilted to positive (negative) $x$ with respect to the beam axis for positive (negative) $\eta_\parallel$. 
The tilted  fireball  
develops a rapidity-odd directed flow $v_1$ of charged particles  in the hydrodynamic expansion 
\cite{Bozek:2010bi,Ryblewski:2012rr,*Konchakovski:2014gda,*Becattini:2015ska},
consistent with experimental data \cite{Abelev:2008jga}.

Charm quarks produced by hard binary collisions between nulceons from the two colliding nuclei are 
forward-backward symmetric in rapidity. The distribution of the production points of charm quarks is 
according to the binary collision profile which is symmetric with respect to the beam axis. However, 
the shift in the bulk and binary collision profiles means that heavy quarks are produced at points in 
the transverse plane shifted with respect to the center of the fireball. This results in the enhanced 
dipole asymmetry in the heavy quark flow pattern. There is a possibility for the heavy quark profile to 
develop a tiny tilt due to correction from nucleon position dependent nuclear modifications of parton 
distribution functions. In this first study we ignore such effects.

The matter in the fireball  is evolved with   3+1 dimensional 
relativistic viscous  hydrodynamics using the HLLE 
algorithm ~\cite{Karpenko:2013wva}. We use a   
shear  viscosity $\eta/s=0.08$ and bulk viscosity $\zeta/s=0.04$ in the hadronic phase. A nonzero bulk 
viscosity in the hadronic phase mimics the  less effective equilibration processes.
 We use a cross-over equation of state, interpolating 
between lattice QCD data  and hadron resonance gas
 model \cite{Bozek:2011ua}.
 Note that the discussed  mechanism of symmetry breaking does not require
 event by event fluctuations 
and all calculations are performed using  smooth initial 
conditions (Eq.~\ref{eq.ic}) from the nucleon 
 Glauber model. A  rapidity-odd directed flow of charged particles of similar 
 magnitude has been predicted in event-by-event simulations as well \cite{Bozek:2011ua}.

The hydrodynamic evolution provides us the spacetime history of the fluid velocity $u^\mu$ and 
temperature $T$. We study the phase space evolution of heavy flavor, particularly charm quarks in 
this background through Langevin dynamics
\beqa
\Delta {\bf r}_i &=& \frac{{\bf p}_i}{E}\Delta t\label{eq.Langevinx}\\
\Delta {\bf p}_i &=& -\gamma {\bf p}_i\Delta t + \rho_i\sqrt{2D\Delta t}\label{eq.Langevinp}
\eeqa
where $\Delta {\bf r}$ and $\Delta {\bf p}$ refer to the updates of the position and momentum vectors of the heavy quark respectively 
in time $\Delta t$. $i=x$, $y$ and $z$ refers to the three 
components in Cartesian coordinates. $\gamma$ and $D$ are the drag and diffusion coefficients respectively that characterize the 
interaction between the heavy quark and the medium. We have assumed a diagonal form for the diffusion matrix here which is often 
employed in the study of heavy quark dynamics related to heavy ion phenomenology~\cite{vanHees:2005wb,Cao:2011et,Scardina:2017ipo}. 
$\rho_i$ is randomly sampled from a normal distribution at every time step such that $\la\rho_i\ra=0$ and $\la\rho_i\rho_j\ra=\delta_{ij}$. 
We work in the post-point realization of this stochastic term and take $D=\gamma E T$, where $E=\sqrt{p^2+m^2}$ is the energy of the heavy 
quark with mass $m$. This ensures the long time approach of the heavy quark phase space distribution to the equilibrium Boltzmann-Juttner 
distribution~\cite{He:2013zua}.

\begin{figure*}
\hskip -6mm
 \includegraphics[scale=0.31]{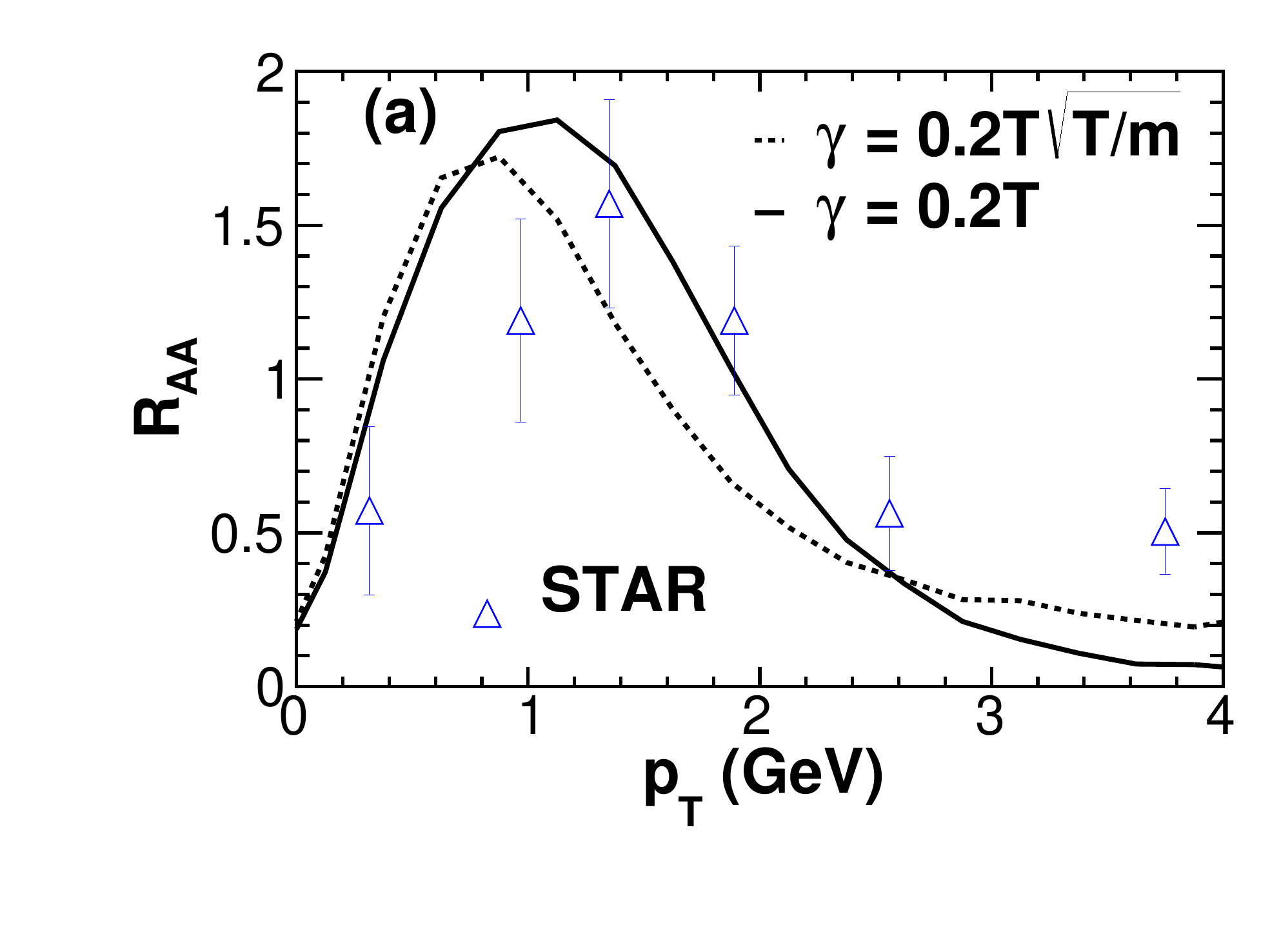}
\hskip -3.mm
 \includegraphics[scale=0.31]{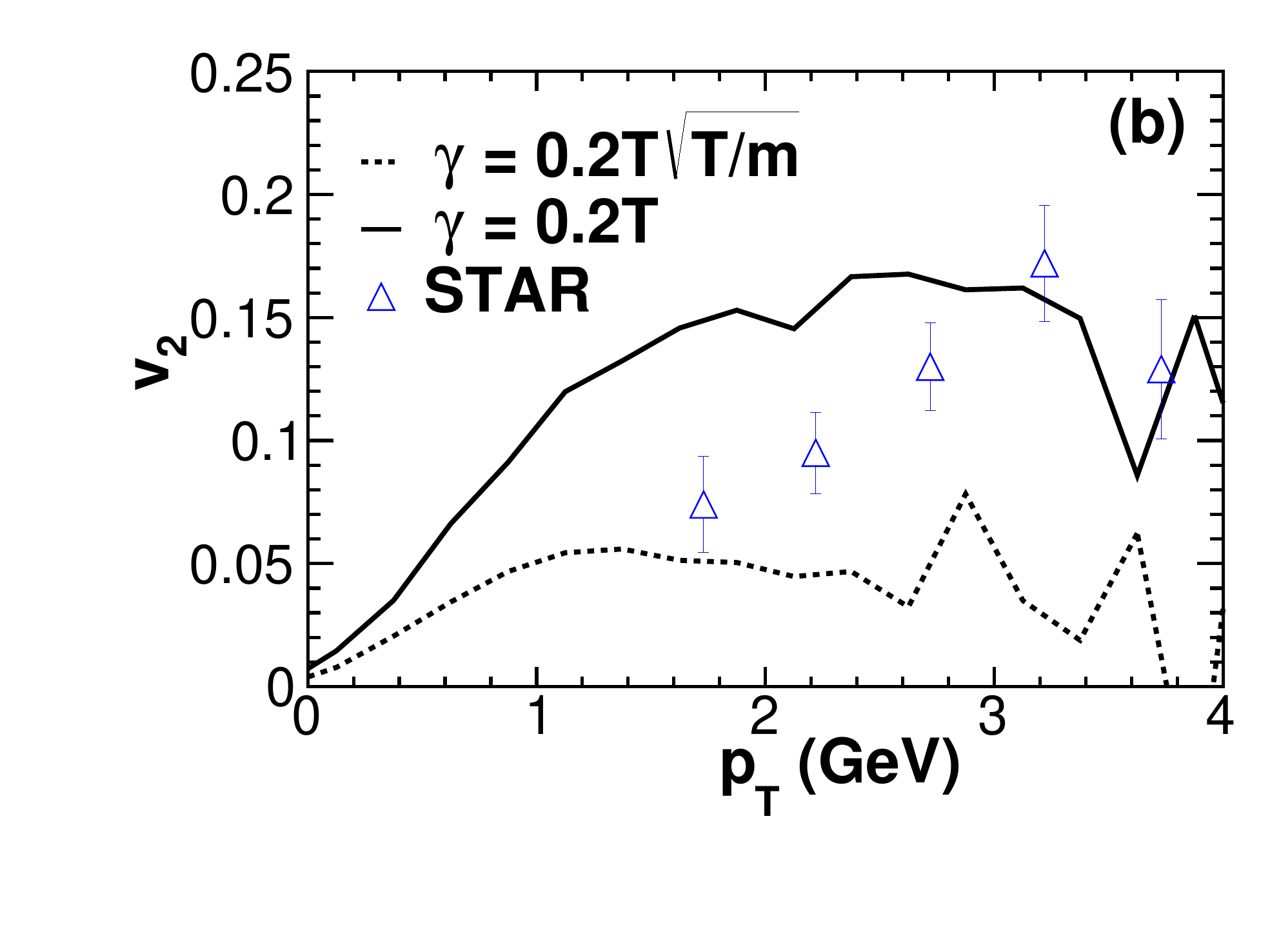}
\hskip -3.mm
 \includegraphics[scale=0.31]{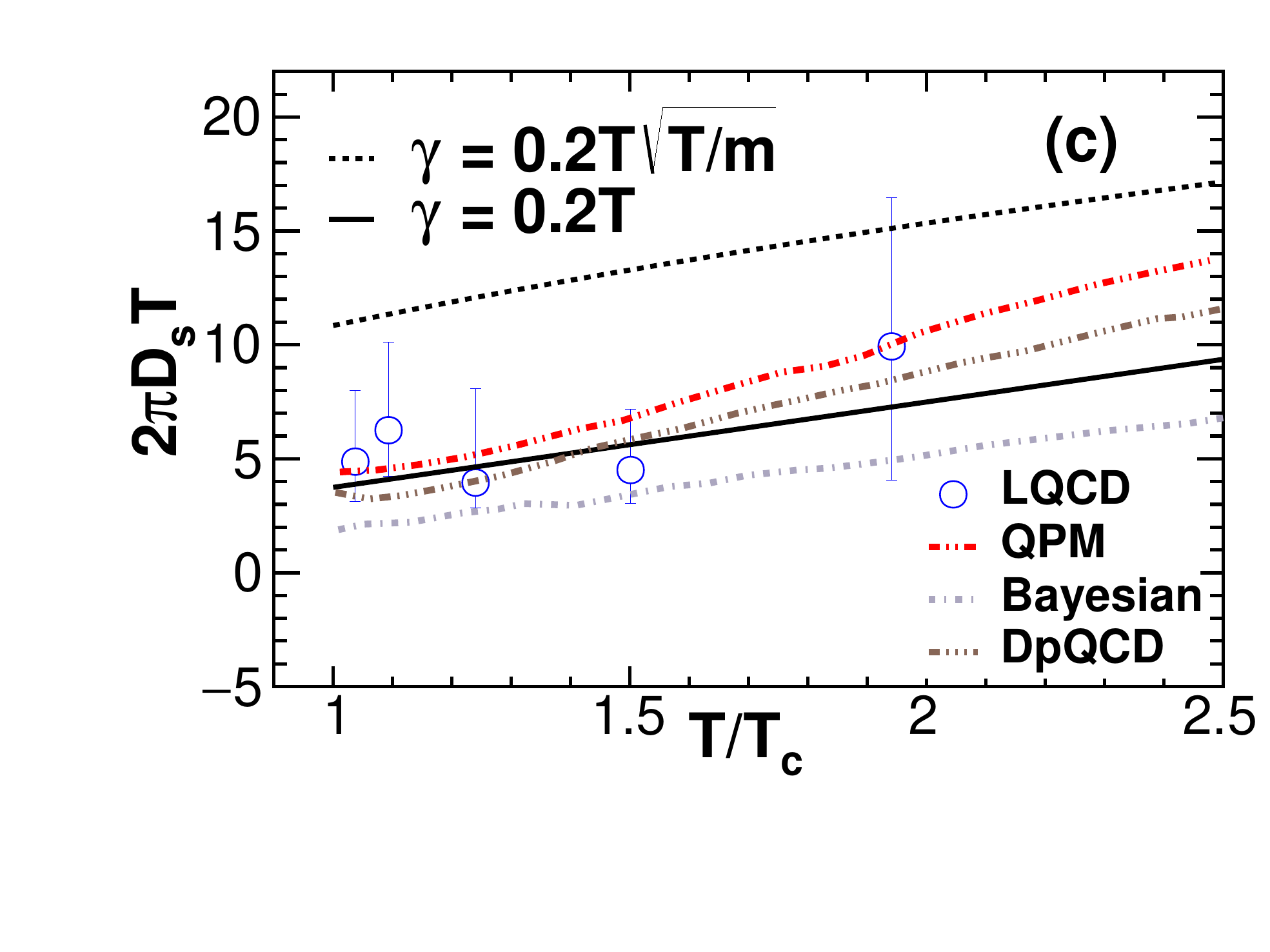} 
\vskip -4mm
 \caption{ (a) Nuclear modification factor for D mesons at 
mid-rapidity as measured in $\l0-10\r\%$ centrality Au+Au 
 at $\sNN=200$ GeV~\cite{Adamczyk:2014uip} compared to 
 our calculation for two different choices for the temperature dependence
of the drag coefficient, 
  $\gamma=0.2T$ and $\gamma=0.2T\sqrt{T/m}$.
 (b) Elliptic flow at mid-rapidity as measured in 
 $\l0-80\r\%$ centrality Au+Au at $\sNN=200$ GeV~\cite{Adamczyk:2017xur}  compared to our calculation for the two 
 different choices of $\gamma$. (c) The spatial diffusion constant for
 the two choices of $\gamma$ in this work compared to results from 
 lattice calculations~\cite{Banerjee:2011ra}, quasiparticle models~\cite{Berrehrah:2014tva,Scardina:2017ipo} and 
 Bayesian analysis of the data~\cite{Xu:2017obm}.}
 \label{fig.raav2}
\vskip -5mm

\end{figure*}

\begin{figure}
 
 \includegraphics[scale=0.35]{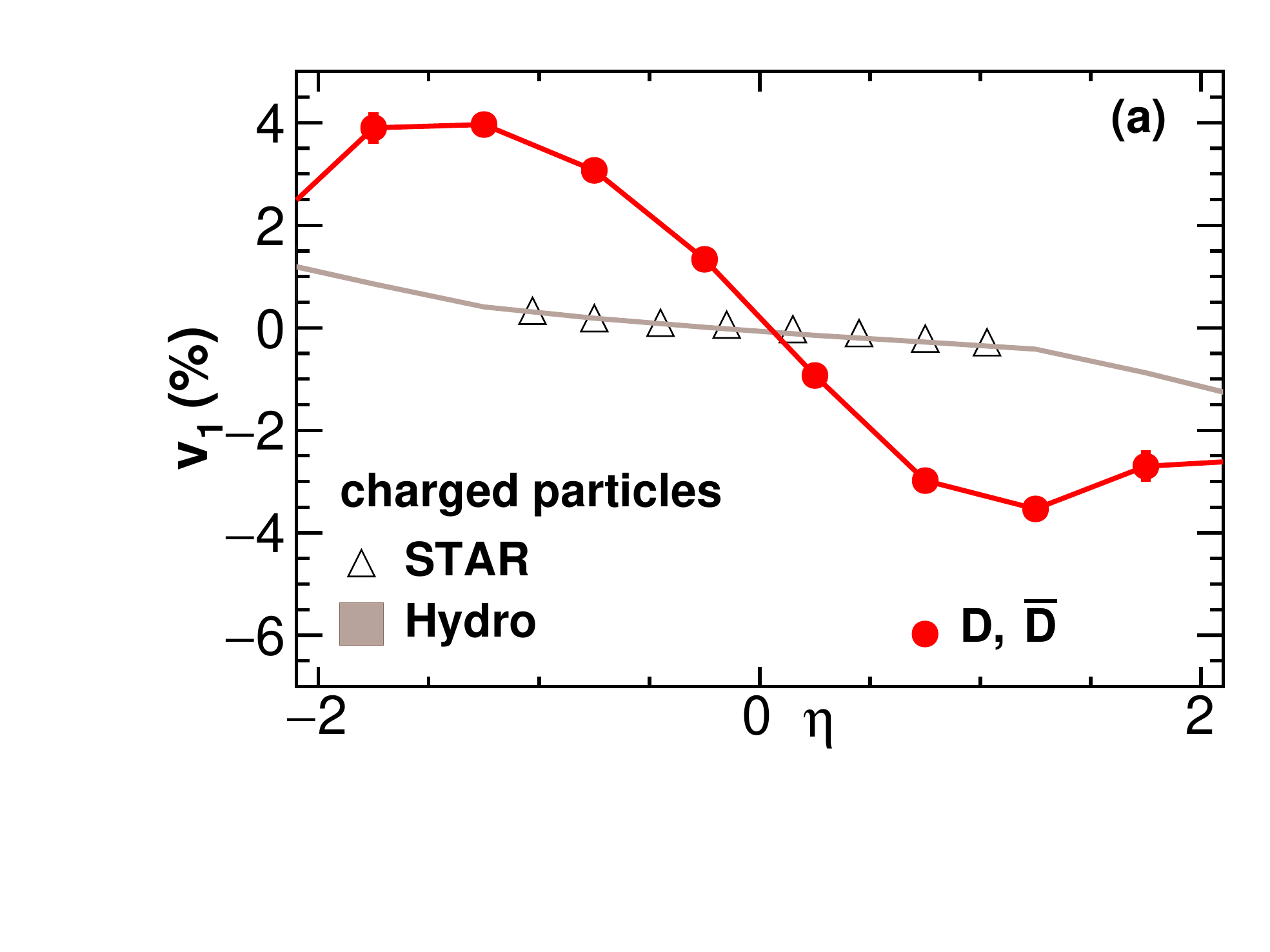}

 \includegraphics[scale=0.35]{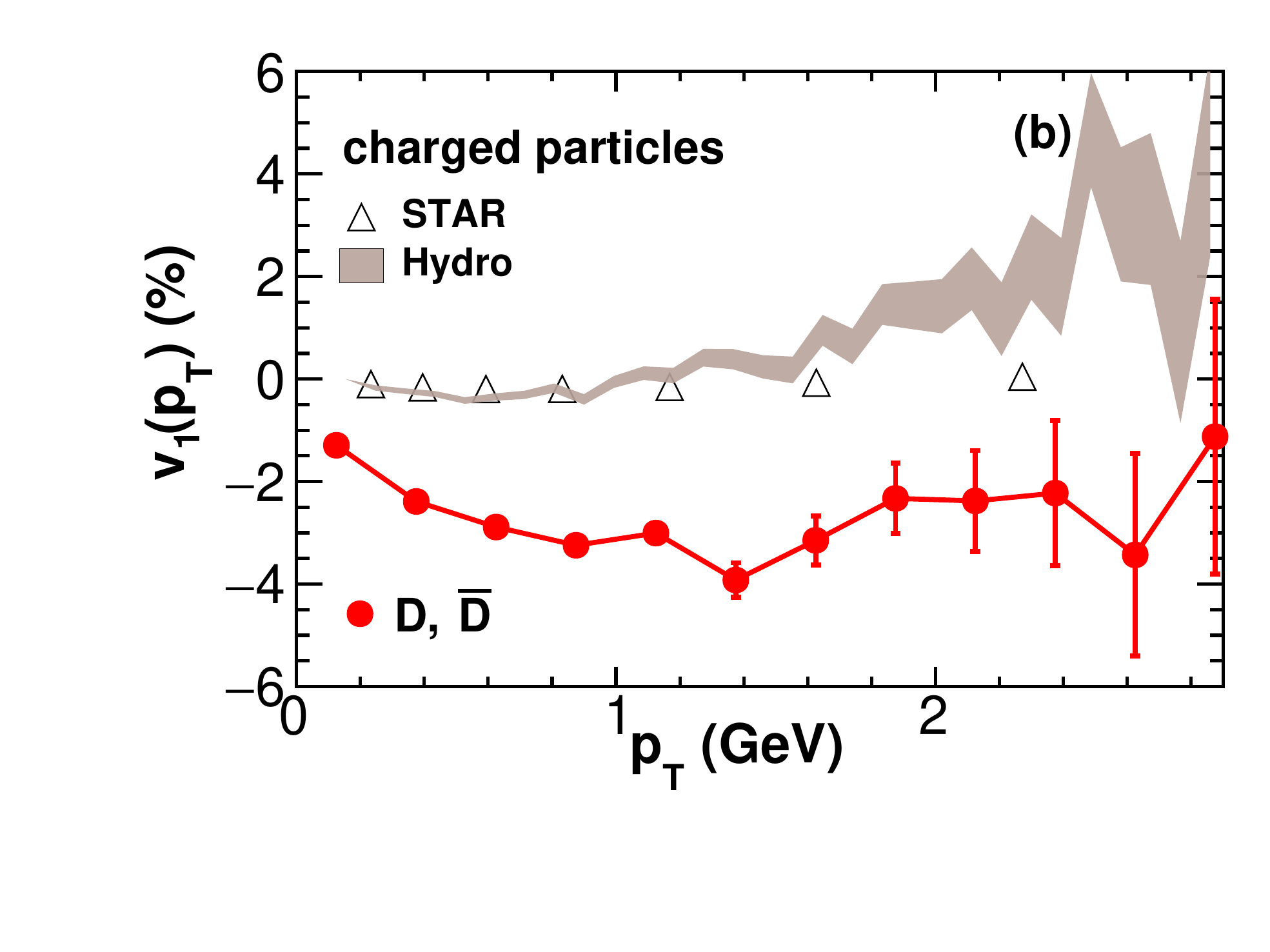} 
\vskip -4mm
 \caption{ (a) Rapidity dependence of $v_1$ for  D mesons as obtained for $\eta_T=3.36$ and $\gamma=0.2 T $ (filled circles). 
 The STAR measurements~\cite{Abelev:2008jga} (open triangles),
 as well as our results from the hydrodynamic model (shaded band)
   for charged particle $v_1$ is also shown for comparison. (b) Same as (a) for $p_T$ differential $v_1$.
 \label{fig.v1}}
\vskip -5mm

\end{figure}

The initial phase space distribution of the heavy quarks is sampled from the binary collision profile and from  the $p_T$ spectra obtained 
in p+p collisions \cite{Cacciari:2005rk,*Das:2013kea}. The Langevin dynamics as written in Eqs.~\ref{eq.Langevinx} and \ref{eq.Langevinp} holds true in a static medium. 
Thus, we perform Lorentz transformation on the heavy quark phase space coordinates to obtain the corresponding coordinates in the 
local rest frame of the fluid following which the position and momentum are updated according to the Langevin evolution of  
Eqs.~\ref{eq.Langevinx} and \ref{eq.Langevinp}. Finally, we perform inverse Lorentz transformation to obtain the heavy quark phase 
space coordinates in the laboratory rest frame. At the end of the Langevin evolution (when the temperature drops to $150$MeV), we hadronize the charm quarks via Petersen 
fragmentation \cite{Peterson:1982ak,*Das:2015ana} to obtain the final momentum distribution of the $D$ and $\bar{D}$ mesons.

The experimental observables for heavy flavor mesons are
 the nuclear suppression factor
\begin{equation}
R_{AA}=\frac{dN^{AA}/dp_T}{N_{coll} dN^{pp}/dp_T} \ ,
\end{equation}
defined as the scaled ratio of $p_T$ spectra in Au+Au and p+p collisions,
and the directed $v_1$ and  elliptic $v_2$ harmonic flow coefficients of 
the azimuthal angle distribution for $D$ mesons
\begin{equation}
\frac{dN}{d\phi} \propto 1 + 2 v_1 \cos(\phi - \Psi_1) + 2 v_2 \cos\left(2 (\phi-\Psi_2)\right) + \dots \ .
\end{equation}
The  elliptic flow is defined with respect to the second order event plane direction $\Psi_2$ for charged particles and the rapidity odd directed flow is defined with respect to the reaction plane $\Psi_1$. In the simulation the reaction plane is well defined, in the experiment it is usually defined from the spectators 
(e.g. see \cite{Steinheimer:2014pfa} for a discussion).

The jury is still out on the determination of the drag experienced by the charm quark in the medium~\cite{vanHees:2004gq, Moore:2004tg, 
Gubser:2006qh,Berrehrah:2014tva,Scardina:2017ipo,Xu:2017obm}. We work with a simple ansatz,
\beqa
\gamma &=& \gamma_0T\l T/m\r^{x}
\label{eq.gamma}
\eeqa
and adjust $\gamma_0$ and $x$ for a qualitative description of the measured values of $R_{AA}\l p_T\r$\cite{Adamczyk:2014uip} and 
$v_2\l p_T\r$\cite{Adamczyk:2017xur} at mid-rapidity.
The results are shown in Fig.~\ref{fig.raav2}. 
A good qualitative description 
of the data is found for $\gamma_0=0.2$ and $x=0-0.5$.
 A different choice, $\gamma = \gamma_0T\l T/E\r^{x}$ does not lead to further 
improvements. In Fig.~\ref{fig.raav2} (c), we have compare the 
$T$ dependence of the dimensionless quantity $2\pi D_sT$ ($D_s$ is the 
charm quark diffusion coefficient in the coordinate space) to
results from  lattice calculations 
~\cite{Banerjee:2011ra}, quasiparticle models~\cite{Berrehrah:2014tva,Scardina:2017ipo} 
and Bayesian analysis of the data~\cite{Xu:2017obm}, lattice 
data slightly favors the choice $\gamma
\propto T$  for the ansatz in Eq.~\ref{eq.gamma}. Nevertheless,  
we study also the case  $\gamma \propto T^{1.5}$ for completeness.
 Different choices do not modify
our conclusions qualitatively.

To make a useful prediction for $v_1$
for experiments with limited statistics we
  show the  0-80 $\%$ centrality results. 
This corresponds to a choice of the impact parameter $b=$~8.3 fm in the 
Glauber model to generate the initial conditions. 
The results shown in Fig. 
\ref{fig.v1} correspond to $\gamma \propto T$. Charged particle spectra 
are calculated  in a statistical emission  model and resonance decays at the  
freeze-out  temperature
$150$~MeV \cite{Chojnacki:2011hb}. The agreement of the directed 
flow coefficient $v_1$
for charged particles with experiment
\cite{Bozek:2010bi} is reproduced in our calculation.

In Fig.~\ref{fig.v1} (a), we show the results for the 
 $p_T$ integrated $v_1$ in different $\eta$ bins. The most notable observation is that for central rapidities
the predicted directed flow coefficient for heavy flavors  is 
{\it  several times larger }
than the $v_1$ of  charged particles   measured by STAR.
It is a very strong, distinctive signal of symmetry breaking in the initial conditions.
Hydrodynamic expansion of the tilted source results in a moderate value of the
odd-component  of the  directed flow for
 charged particles stemming from the thermalized fireball.
 The heavy quark production points are shifted  in the transverse plane with 
respect to the bulk of the matter. The interaction of the medium 
 in the fireball on heavy quarks is biased resulting
 in a larger directed flow of heavy flavor mesons.
We note that our prediction for heavy flavor $v_1$ at central rapidities 
is approximately three times
 larger than in the results of earlier calculations using a transport
 model \cite{Bratkovskaya:2004ec}. 
 
Initially, the bulk of  matter is tilted which 
translates into  a tilted temperature profile. This sets off
 the hydrodynamic response and the directed flow of the fluid
 is built after some 
time \cite{Bozek:2010bi}. Thus, there could be two factors that drive the large $v_1$ of heavy flavors,  preferential scattering of the heavy quarks by 
the tilted source  and secondly, the drag by the directed flow of the
 bulk matter. 
We have explicitly investigated this by using boost invariant $T$ or $u^{\mu}$ 
and found that the large heavy flavor $v_1$ is mostly due to the drag 
of the fluid.

In Fig.~\ref{fig.v1} (b), we show the $p_T$ differential $sgn(\eta) \ v_1$ 
in the $|\eta|\leq 1.3$ range.  The $p_T$ dependence of $v_1$ is similar to that of $v_2$ from Fig.~\ref{fig.raav2} (b): it rises with $p_T$ 
for $p_T<1$ GeV and then stays constant up to $p_T\sim2$ GeV before slowly falling off. The $p_T$ range where
 the directed flow magnitude is large   is also the $p_T$ range where
 $R_{AA}>1$ as seen 
in Fig.~\ref{fig.raav2} (a). The relative modification of the $p_T$ for heavy quarks in this momentum  range due to a drag from the collective motion
of matter is the largest, and it is at the origin of 
both the increase of $R_{AA}$
and a large  $v_1$. At higher $p_T$, jet quenching by the tilted fireball~\cite{Adil:2005qn} could be a source of additional $v_1$ besides the 
drag by the tilted bulk matter that we discuss.
The spectra of heavy flavor mesons are shifted towards negative
$p_x$.  The bias from the off-center
position of the fireball (Fig. \ref{fig.tilt}) leads to 
a net transfer of $p_x$ to heavy quarks. 

\begin{figure}
 \begin{center}
 \includegraphics[scale=0.35]{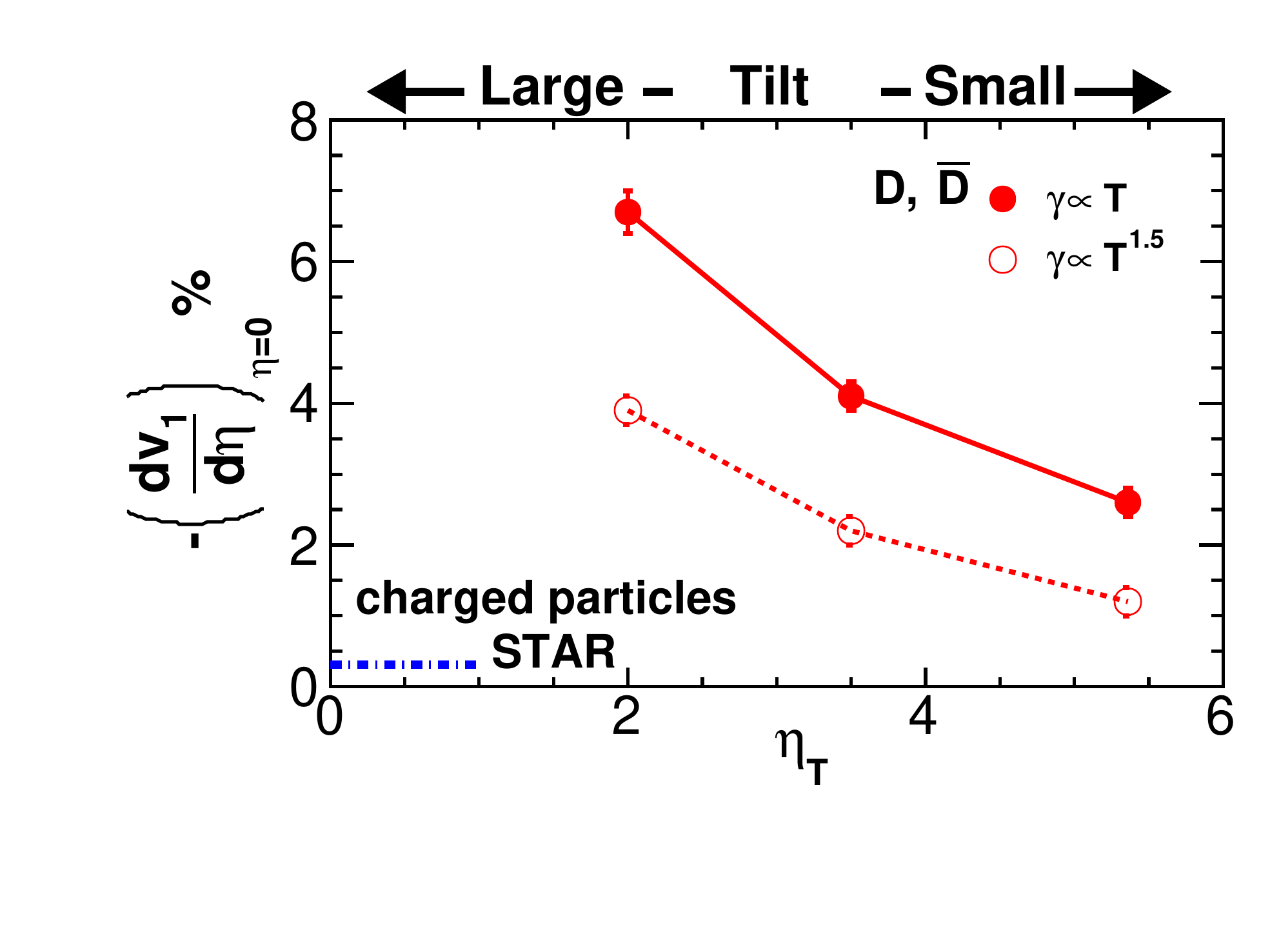}
\vskip -9mm
 \caption{Slope of the directed flow at mid-rapidity for D and $\bar{D}$ mesons compared to that of the STAR measurement of mid-rapidity 
slope for  charged particles~\cite{Abelev:2008jga}.
}
 \label{fig.slope}
\vskip -5mm

 \end{center}
\end{figure}

The observation of enhanced forward-backward dipole asymmetry 
in the flow of heavy quarks is best depicted by the $v_1$ slope at 
mid-rapidity. This is shown in Fig.~\ref{fig.slope}. The slope has 
been extracted by fitting to the $v_1$ in the $|\eta|<1$ range. Depending
 on the choice of the parameters the  
slope for the D mesons is $5-20$ times larger than that
 of the charged particles.
 A notable feature of the directed flow of heavy 
flavors is its sensitivity to the magnitude of the initial tilt of the fireball. 
By varying $\eta_T$ governing the asymmetry of the initial
 entropy deposition of the left and right going sources (Eq. \ref{eq.fpm}) 
 the initial tilt can be varied. When changing $\eta_T$ from 
$y_{beam}$ (small tilt) to $0.4y_{beam}$ (large tilt), 
the change in charged particle $v_1$ is moderate, $\sim0.5\%$ at $\eta=0.75$.
On the other hand, the heavy flavor $v_1$ at the same $\eta$ bin changes 
by $2-3\%$ while the $v_1$ slope increases 
by a factor $3$. Thus, a measurement of the  $v_1$
 slope for heavy flavors
 would present a sensitive probe of the initial matter distribution 
in the transverse and longitudinal directions.

Several comments are in order. The dependence of $\gamma$ on the medium properties like $T$ and heavy quark properties like $m$ and $p$ is 
far from settled~\cite{vanHees:2004gq, Moore:2004tg,Gubser:2006qh,Berrehrah:2014tva,Scardina:2017ipo}. The value of the drag coefficient 
influences the final $v_1$ significantly and hence ignorance of $\gamma$ can 
introduce considerable uncertainty in the extraction of $\eta_T$. Thus, proper calibration of the medium interaction with the heavy quark and 
its demonstration by good simultaneous description of $p_T$ dependence of $R_{AA}$ and $v_2$ at mid-rapidity is essential.
 The directed flow of heavy flavors should  be
included in the set of observables used to constrain the 
 drag coefficient, besides $R_{AA}$ and $v_2$.

We have checked 
that the heavy quark $v_1$ is
 robust to variations of other parameters like the initial time to start hydrodynamics $\tau_0$, the value of shear and bulk viscosity $\eta/s$ and $\zeta/s$. 
It has been suggested that hard processes in a $k_T$ 
factorized saturation framework follow a profile that is also tilted with respect to the beam axis~\cite{Adil:2005bb}. This would reduce  the heavy 
quark $v_1$ or could even reverse its sign.
 Thus, the relative magnitude (and sign)
 of heavy flavor  and charged particle $v_1$ can distinguish between different 
saturation mechanism for hard processes. 
This emphasizes the significance of the measurement of heavy quark $v_1$ at different collision energies.

The strong electromagnetic  fields in the initial state could contribute
 to heavy flavor $v_1$ as 
well~\cite{Das:2016cwd}. However, the effect of the electromagnetic fields is of
 opposite sign on $D$ and $\bar{D}$ mesons and would not influence the average
directed flow of  $D$ mesons.

We note that at higher energies the magnitude of the rapidity-odd 
directed flow of charged particles decreases. In  the hydrodynamic model 
it results from a decrease of the  initial tilt of the source. This effect
 would lead to a decrease of the directed flow of  $D$ mesons at higher energies. 
On the
 other hand, the larger density of the fireball makes the
 drag of heavy quark more effective.  The  resulting  $D$ meson directed 
flow at $\sqrt{s_{NN}}=2760$~GeV is predicted to be 
 much larger than for charged particles.

The authors acknowledge helpful discussion with Iurii Karpenko on the vHLLE code. SC also acknowledges fruitful discussions with Jane Alam, Santosh K. Das, 
Vincenzo Greco and Suropriya Saha. 
This research is supported by the AGH UST statutory tasks
No. 11.11.220.01/1 within subsidy of the Ministry of Science and
Higher Educations and by the National Science Centre Grant
No. 2015/17/B/ST2/00101.

\bibliographystyle{apsrev4-1-nohep}
\bibliography{FBHQ}

\end{document}